\documentclass[aps, prb, reprint, superscriptaddress, preprintnumbers,amsmath,amssymb,twocolumn,floatfix]{revtex4-2}
\usepackage{amssymb}
\usepackage{amsmath}
\usepackage{bm}
\usepackage[dvips]{graphicx}
\usepackage{dcolumn}
\usepackage{longtable}
\usepackage{color}
\usepackage[version = 4]{mhchem}
\usepackage{comment}

\newcommand{\subm}[1]{_{\mathrm{#1}}}  
\newcommand{\LRA}{LaRh$_2$As$_2$~}
\newcommand{\CRA}{CeRh$_2$As$_2$~}
\newcommand{\Tc}{$T\subm{c}$}

\begin{document}
\title{Conventional $s$-wave Superconductivity in LaRh$_2$As$_2$; \\the Analog without the 4$f$ Electrons of CeRh$_2$As$_2$}

\author{Shiki \surname{Ogata}}
\email{ogata.shiki.86c@st.kyoto-u.ac.jp}
\affiliation{Department of Physics, Kyoto University, Kyoto 606-8502, Japan}
\author{Shunsaku \surname{Kitagawa}}
\affiliation{Department of Physics, Kyoto University, Kyoto 606-8502, Japan}
\author{Kenji \surname{Ishida}}
\affiliation{Department of Physics, Kyoto University, Kyoto 606-8502, Japan}

\author{Manuel \surname{Brando}}
\affiliation{Max Planck Institute for Chemical Physics of Solids, D-01187 Dresden, Germany}

\author{Elena \surname{Hassinger}}
\affiliation{Institute for Quantum Materials and Technology, Karlsruhe Institute of Technology, 76131 Karlsruhe, Germany}
\affiliation{Max Planck Institute for Chemical Physics of Solids, D-01187 Dresden, Germany}

\author{Christoph \surname{Geibel}}
\affiliation{Max Planck Institute for Chemical Physics of Solids, D-01187 Dresden, Germany}
\author{Seunghyun \surname{Khim}}
\affiliation{Max Planck Institute for Chemical Physics of Solids, D-01187 Dresden, Germany}
\date{\today}

\begin{abstract}
Superconductor \LRA has the same crystal structures as CeRh$_2$As$_2$, which exhibits superconducting (SC) multiphase in the $c$-axis magnetic field. Although the SC transition temperatures \Tc \ are similar, around 0.3 K, \LRA shows conventional type-II superconductivity with a small upper critical field $H\subm{c2}\sim$ 10 mT. At present, the SC properties of \LRA have not been clarified yet. We performed $^{75}$As-nuclear quadrupole resonance (NQR) measurements on \LRA to investigate the SC properties and gap structure. $1/T_1$ shows a clear coherence peak just below \Tc \ and an exponential decrease at lower temperatures, suggesting full-gap $s$-wave superconductivity. The numerical calculations based on an $s$-wave SC model reveal an SC gap size of $\Delta(0)/k\subm{B}T\subm{c} \sim 1.48$, consistent with the weak-coupling $s$-wave superconductivity. These results suggest that the 4$f$ electrons in \CRA not only enhance the orbital limiting field but also contribute to the formation of unconventional superconductivity with SC multiphase.
\end{abstract} 

\maketitle

In inversion symmetry breaking systems, antisymmetric spin-orbit coupling (SOC) causes spin splitting of the Fermi surface \cite{RashbaASOC, DresselhausASOC,CeIrSi3_dHv,Yb4Sb3_Kimura,JPSJ_Cr3P}. Antisymmetric SOC and the resulting spin splitting of the Fermi surface can lead to a unique superconducting (SC) state. In globally inversion-symmetry breaking superconductors, antisymmetric SOC induces parity mixing, which results in the mixing of spin-singlet and spin-triplet superconductivity and an enhancement of $H\subm{c2}$ \cite{ASOCSC,CePt3Si_theory,CePt3SI_NMR,CePt3Si_NMR_theory,CeIrSi3_NMR}. In locally inversion-symmetry breaking superconductors, alternating antisymmetric SOC can give rise to field-induced SC multiphase, accompanied with parity transition from even to odd parity SC state \cite{T_Yoshida}.\par
The heavy-fermion superconductor \CRA with a SC transition temperature $T\subm{c} \sim$ 0.3 K shows the SC multiphase under the $c$-axis magnetic field \cite{hakken_CRA}. \CRA crystallizes in a locally-inversion symmetry-breaking tetragonal CaBe$_2$Ge$_2$-type structure with the space group $P4/nmm$ (No. 129, $D^{7}\subm{4h}$), which has two Ce sites in the unit cell with opposite Rashba SOC. Such a staggered Rashba SOC is considered to be attributed to the SC multiphase. In addition, \CRA also exhibits phase transitions above and below $T\subm{SC}$ ($T_{0} \sim$ 0.4 K and $T\subm{N} \sim$ 0.25 K) \cite{hakken_CRA, QDW_CRA, QDW_theta_CRA, QDW_new_CRA, QDW_Los_CRA, Kibune,H||c_ogata,H||ab_ogata,Khanenko_phasediagram}. The order parameter below $T_0$ remains a subject of debate \cite{QDW_CRA,T0_AFM_theory,Miyake_T0}. \CRA is an intriguing material in which superconductivity, magnetism, and the other phase ($T_0$) are intertwined.\par
LaRh$_2$As$_2$, which has the same crystal structure but no 4$f$ electrons, also exhibits superconductivity with the similar transition temperature \Tc $\sim$ 0.3 K \cite{LRA_MPI}. Despite having the same crystal structure and a similar SC transition temperature, \LRA does not exhibit a field-induced SC multiphase \cite{LRA_MPI}. The upper critical fields $H\subm{c2}$ of \CRA is $\sim$14 T for the $c$-axis field and $\sim$2 T for the $ab$-plane field, while those of \LRA are about 10 mT in both directions. Although it is considered that the 4$f$ electrons play a crucial role in the unconventional SC properties of CeRh$_2$As$_2$, it remains unclear whether the absence of the field-induced SC multiphase in \LRA is simply due to the small orbital-limited field, or the presence of the additional 4$f$-electron interactions connected with the SC multiphase. Although Numerical calculations suggest that \LRA is an electron-phonon-mediated superconductor in the weak-coupling limit \cite{LRA_MPI}, experimental evidence for its SC gap structure has not been reported. Therefore, experiments to detect the quasiparticle density of states (DOS) in the SC state are required.\par
In this paper, we report the $^{75}$As-nuclear magnetic resonance (NMR) and NQR results of \LRA to investigate the SC order parameter. We measured nuclear spin-lattice relaxation rate $1/T_1$ at the As(2) site, one of the two crystallographically inequivalent As sites. $1/T_1$ shows a clear coherence peak just below $T\subm{c}$ and an exponential decrease at lower temperatures, indicating full-gap superconductivity. Comparison between numerical calculation of $1/T_1$ and the present results suggests a SC gap size of $\Delta/k\subm{B}T\subm{c} \sim 1.48$, which is consistent with the weak-couple $s$-wave superconductivity. Our results provide important clues to discuss the origin of the realization of a field-induced odd-parity spin-singlet SC state due to the local inversion-symmetry breaking.\par
Single crystals of \LRA with a typical size of 2.0 $\times$ 1.5 $\times$ 0.8 mm$^3$ were grown using the bismuth flux method \cite{LRA_MPI}. The SC transition temperature \Tc \  was determined from the SC diamagnetic signal with the ac magnetic susceptibility measurements using an NQR coil. For the NMR measurements to estimate the NQR coupling constant $\nu\subm{Q}$, we used a split SC magnet that
generates a horizontal field and combined it with a single-axis
rotator to control the magnetic field angle precisely. NQR measurements were performed using a $^3$He-$^4$He dilution refrigerator, in which the sample was immersed into the $^3$He-$^4$He mixture to reduce radio-frequency heating during measurements. A conventional spin-echo method was used for the NMR and NQR measurements. The nuclear spin-lattice relaxation rate 1/$T_1$ was determined by fitting the time variation of the nuclear magnetization probed with the spin-echo intensity after saturation to a theoretical function for $I = 3/2$ \cite{T1function1,T1function2}.\par
%
%
\begin{figure*}[htbp!]
   \begin{center}
   \includegraphics[width=\linewidth]{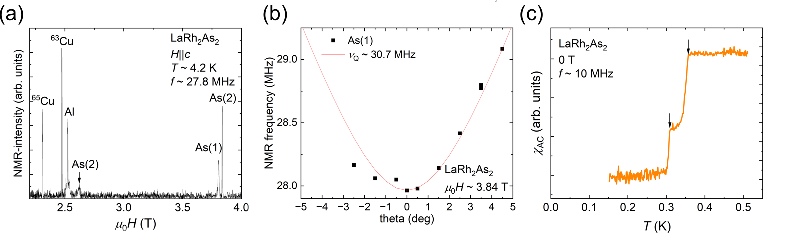}
   \end{center}
   \caption{(Color online) (a) The NMR spectrum of \LRA at 4.2 K in $H\parallel c$. (b) The field angular dependence of the central peak of the As(1) site at 3.84 T. (c) Temperature dependence of ac magnetic susceptibility $\chi\subm{AC}$ measured at the frequency of 10 MHz under zero field. The black arrows indicate the two-step superconducting transition temperatures.}
\end{figure*}
We performed NMR measurements in the $c$-axis magnetic field to estimate the NQR coupling constant $\nu\subm{Q}$ of the As sites, as shown in Fig.~1 (a). In the nuclei with nuclear spin $I\geq 1$, the degeneracy of the nuclear energy levels is lifted by the Zeeman ($\mathcal{H}\subm{Z}$) and the electric quadrupole ($\mathcal{H}\subm{Q}$) interactions. The total nuclear Hamiltonian $\mathcal{H}$ is described as
\begin{flalign}
    \mathcal{H} =&\mathcal{H}\subm{Z} + \mathcal{H}\subm{Q} \nonumber\\
    = &-\gamma \hbar (1+K_i)I\cdot H \nonumber\\& + \frac{h\nu\subm{Q}}{6}\left[3I^{2}_{z} - I(I+1) +\frac{\eta}{2}(I^{2}_{+} + I^{2}_{-})\right],
\end{flalign}
where $K_i$ ($i=c$ and $\perp$) is the Knight shift along the $c$ axis and perpendicular to the $c$ axis, $h\nu\subm{Q} = \{3eQV_{zz}/[2I(2I-1)]\}$ is the NQR coupling constant, $\eta = |(V_{yy} - V_{xx})/V_{zz}|$ is the asymmetric parameter, and $V_{ii}$ is the electric field gradient (EFG) along the $i$ axis ($i = x, y, z$). The $z$ axis is defined as the principal axis of the EFG tensor with the largest eigenvalue. In LaRh$_2$As$_2$, the $z$ axis is the $c$ axis and $\eta$ is zero at each As site because of the fourfold symmetry of the atomic position.\par
The two peaks around 3.8 T are the central peaks (1/2 $\leftrightarrow$ -1/2) of the two As sites. In addition to the $^{27}$Al, $^{63}$Cu, $^{65}$Cu signals around 2.5 T, there is a rather weak peak, which is a satellite peak of As(2). The difference from the central peak is $\sim$1.216 T, from which $\nu\subm{Q}$ can be estimated as $\nu\subm{Q} \sim$ 8.86 MHz. The satellite peak of As(1) could not be observed in this setup, so we measured the field angular dependence of central peak of As(1) to estimate the $\nu\subm{Q}$, as shown in Fig.~1 (b). From the field angular dependence, the $\nu\subm{Q}$ is estimated to be 29.7 MHz. Comparing with the $\nu\subm{Q}$ in CeRh$_2$As$_2$ [As(1): 31.1 MHz, As(2): 10.75 MHz], we assign the site with $\nu\subm{Q} \sim$ 8.86 MHz as the As(2) site and the site with $\nu\subm{Q} \sim$ 29.7 MHz as the As(1) site, respectively.\par
\begin{figure}[htbp!]
   \begin{center}
   \includegraphics[width=0.98\linewidth]{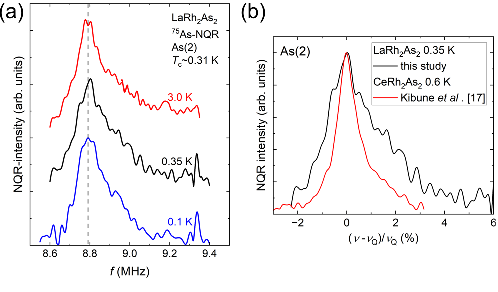}
   \end{center}
   \caption{(Color online) (a) Temperature evolution of the NQR spectra of the As(2) site. (b) The NQR frequency distribution $(\nu - \nu\subm{Q})/\nu\subm{Q}$ at the As(2) site of \LRA and \CRA \cite{Kibune}.}
\end{figure}
Figure 1(c) shows the results of the ac susceptibility measurements using the NQR coil. SC diamagnetic signal was observed below 0.36 K. Here, the ac susceptibility shows a two-step anomaly, likely due to the inhomogeneity of the sample. The SC transition temperatures were determined to be $T\subm{c1} \sim$ 0.36 K and $T\subm{c2} \sim$ 0.31 K. \par
Figure 2(a) shows the $^{75}$As-NQR spectra at the As(2) site at various temperatures, showing clear peaks around 8.8 MHz. The peak frequency is almost temperature independent up to 3.0 K. The NQR signal at the As(1) site could not be observed in our setup.  
A comparison between the NQR spectra in \LRA and those in CeRh$_2$As$_2$ in Fig.~2(b) suggests that the sample quality in \LRA would be not as good as that of \CRA in ref.~\cite{Kibune}. The NQR frequency distribution $\Delta \nu\subm{Q}/\nu\subm{Q}$ at the As(2) site is around 1.1 \% for CeRh$_2$As$_2$, whereas it is around 2.3 \% for LaRh$_2$As$_2$. In addition, the NQR spectrum of \LRA has a more noticeable shoulder on the high-frequency side than that observed in the As(2)-site NQR of \CRA \cite{Kibune}, as seen in Fig.~2(b). The shoulder structure becomes less pronounced in the recent higher-quality \CRA sample. This suggests that the shoulder originates from the inhomogeneity in the sample, which is probably caused by the stacking faults along the $c$ axis.\par
\begin{figure}[htbp!]
   \begin{center}
   \includegraphics[width=0.95\linewidth]{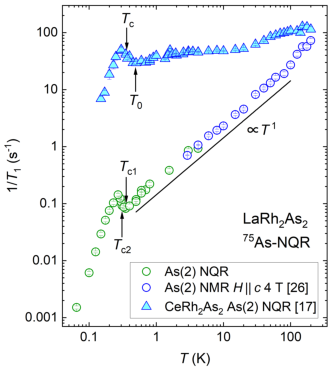}
   \end{center}
   \caption{(Color online) Temperature dependence of 1/$T_1$ measured at the peak frequencies of the As(2) site. The blue and green circles represent the results of the $c$-axis field NMR \cite{Kibune_normal} and zero-field NQR measurements, respectively. The light blue triangles represent the NQR 1/$T_1$ of the As(2) site of CeRh$_2$As$_2$ as references \cite{Kibune}. The arrows indicate $T\subm{c}$s and $T\subm{0}$.}
\end{figure}
To clarify the SC gap structure, we measured $1/T_1$ of the As(2) site, which reflects the quasiparticle DOS near the Fermi energy. Figure 3 shows the temperature dependence of $1/T_1$ of the NQR at the As(2) site. For reference, we also show the results measured with NMR under $H\ ||\ c$ on LaRh$_2$As$_2$, together with the results of the NQR on CeRh$_2$As$_2$ \cite{Kibune_normal,Kibune}. As seen in Fig.~3, $1/T_1$ agrees with the NQR and NMR ($H\ ||\ c, \mu_{0}H\sim 4$ T) results at 4.2 K and 1.5 K. This indicates that the electronic state is field-independent up to 4 T. Here, it is noted that the fluctuations detected by the NMR $1/T_1$ and NQR $1/T_1$ are the same direction, since the EFG principal axis is the $c$ axis.\par
The temperature dependence of $1/T_1$ in the normal state of \LRA is approximately linear to temperature in the entire measured temperature range. This is typical behavior for normal metals. The Knight shift is also nearly temperature-independent ($K \sim 0.3 \%$) \cite{Kibune_normal}, and $K(\alpha)$ in the Korringa's law $\frac{1}{T_{1}TK^2}=\frac{4\pi k\subm{B}}{\hbar}\left(\frac{\gamma_n}{\gamma_e}\right)^2K(\alpha)$ is approximately 0.3, independent of temperature. 
Although $K(\alpha)$ is less than 1, 1/$T_1T$ is almost temperature independent, suggesting that it is due to electron-electron interactions rather than magnetic fluctuations. A $K(\alpha)$ values less than 1 due to electron-electron interactions are also observed in simple metals \cite{Moriya,Narath1,Narath2}. In the SC state, $1/T_1$ shows a pronounced coherence peak below \Tc \ and an exponential decrease at lower temperatures. These behaviors in the SC state strongly suggest that \LRA is a full-gap $s$-wave superconductor.\par
\begin{figure}[htbp!]
   \begin{center}
   \includegraphics[width=0.8\linewidth]{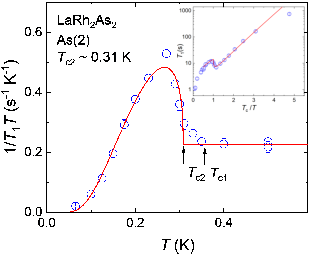}
   \end{center}
   \caption{(Color online) The 1/$T_{1}T$ at the As(2) site plotted against $T$. The calculation using BCS theory is also shown in the red solid curve [Eq. (2)]. The black arrows indicate the SC transition temperatures. The inset shows Arrhenius plot of $T_1$ against \Tc /$T$ (\Tc = 0.31 K).}
\end{figure}
%
%
To evaluate the SC gap size from the temperature dependence of $1/T_{1}$, we reproduced the 1/$T_{1}$ results with a numerical calculation based on BCS theory as shown in Fig.~4. 1/$T_{1}$ in the SC state (1/$T\subm{1s}$) normalized by that in the normal state (1/$T\subm{1n}$) is expressed as 
\begin{flalign}
    \frac{T\subm{1n}}{T\subm{1s}} = \frac{2}{k\subm{B}T}\int^{\infty}_{0}dEN^{2}_{s}(E)&\left[1 + \frac{|\Delta (T)|^2}{E^2}\right]\nonumber \\
    &\times f(E)[1-f(E)],
\end{flalign}
where $N_{s}(E)$ is the quasiparticle DOS in the SC state, $\Delta (T)$ is the temperature dependent SC gap, and $f(E)$ is the Fermi distribution function. The factor $\left[1 + \frac{|\Delta (T)|^2}{E^2}\right]$ is related to the coherence effect in the SC state. We considered the energy broadening in $N_s (E)$ by taking the convolution of $N_s (E)$ with a rectangular broadening function, whose width and height are $2\delta$ and $1/2\delta$, respectively \cite {T1_namarase}. The experimental results are well reproduced using $\Delta(0)/k\subm{B}T\subm{c} = 1.48$ and $\delta/\Delta(0) = 0.088$, as shown in the solid red line in Fig.~4. This $\Delta (0)/k\subm{B}T\subm{c}$ value is close to the weakly coupled $s$-wave BCS theory value of 1.76. This agreement, the existence of coherence peak, and the exponential decay of $1/T_1$ indicate that a conventional full-gap superconductivity is realized in LaRh$_2$As$_2$. 1/$T_1T$ started to increase gradually around $T\subm{c1} \sim$ 0.36 K and increased sharply around $T\subm{c2}\sim$ 0.31 K. Hence, we used the points below $T\subm{SC} = T\subm{c2}\sim$ 0.31 K for the fitting. When tentatively adopting $T\subm{SC} = T\subm{c1}\sim$ 0.36 K, the observed $1/T_1T$ could not be fitted well with the BCS model with a single SC gap. Thus, it is considered that only a part of the sample shows superconductivity in $T\subm{c2}<T<T\subm{c1}$ \par
Finally, we compare our present experimental results with those of \CRA to clarify the strong correlation effect of the 4$f$ electrons. The $1/T_{1}T$ values of CeRh$_2$As$_2$ we have previously measured \cite{Kibune} are about 1000 times larger than those of LaRh$_2$As$_2$, indicating large magnetic fluctuations arising from the Ce 4$f$ electrons. $1/T_1 T$ of \CRA starts to increase on cooling from $T_0$, showing a peak below \Tc , and decreases on further cooling. Such behaviors of $1/T_1 T$ around \Tc \  seems to suggest that superconductivity is driven by AFM fluctuations \cite{BaFe2As2_NMR_AFMfluctuation,KdopeBaFe2As2_NMR_T1T2,BaFe2As2_NMR_AFMfluctuation_pressure,CeRhIn5_NMR_AFMfluctuation,Rb2Cr3As3_NMR_AFMfluctuation}. However, it is noted that in CeRh$_2$As$_2$, the existence of the AFM transition temperature $T\subm{N}$ at the vicinity of \Tc \ makes it difficult to distinguish whether the critical slowing down of $1/T_1T$ in \CRA is caused by superconductivity or AFM transition.\par
Recent micro-Hall probe measurement suggests full-gap superconductivity in \CRA based on the temperature dependence of $H\subm{c1}$ \cite{Hc1_Hall}. However, superconductivity coexists with antiferromagnetism in CeRh$_2$As$_2$ \cite{Kibune,H||c_ogata,H||ab_ogata,muSR_Khim}, ruling out simple BCS-type $s$-wave superconductivity, which is incompatible with magnetism. Interband pairing, as in CeCu$_2$Si$_2$ \cite{fullgapCeCu2Si2_Kittaka,fullgapCeCu2Si2_review_exp,fullgapCeCu2Si2_theory}, may need to be considered. Hence, the conventional $s$-wave superconductivity in \LRA indicates that the superconductivity in LaRh$_2$As$_2$ is essentially different from that in CeRh$_2$As$_2$, even though they have similar SC transition temperatures. This suggests the importance of the electron-electron interactions originating from the 4$f$ electrons for superconductivity in CeRh$_2$As$_2$, and possibly also for the field-induced odd-parity SC state.\par
In conclusion, we performed $^{75}$As-NQR measurements at the As(2) site on the locally non-centrosymmetric superconductor \LRA single crystal to investigate the SC gap symmetry. The temperature dependence of $1/T_1$ exhibits almost Korringa behavior above \Tc , suggesting a conventional metallic behavior in the normal state. It also shows a clear coherence peak just below \Tc \ and an exponential decrease at lower temperatures, providing strong evidence for full-gap $s$-wave superconductivity. The numerical calculation evaluates the SC gap size to be $\Delta(0)/k\subm{B}T\subm{c}\sim 1.48$, which is close to the value of the weakly coupled BCS theory. This suggests that \LRA and \CRA exhibit fundamentally different SC characteristics, even though they have the same crystal structure and similar SC transition temperatures. Our findings provide an important clue for understanding the origin of the SC multiphase arising from the local inversion symmetry breaking.\par
\vskip.5\baselineskip
\begin{acknowledgments}
This work was partially supported by the Kyoto University LTM Center and Grants-in-Aid for Scientific Research (KAKENHI) (Grants No. JP20KK0061, No. JP20H00130, No. JP21K18600, No. JP22H04933, No. JP22H01168, No. JP23K22439, No. JP23H01124, No. JP23K25821, No. JP24KJ1360, and No. JP25H00609). This work was also supported by JST SPRING (grant number JPMJSP2110) and research support funding from the Kyoto University Foundation, and ISHIZUE 2024 of Kyoto University Research Department Program, and Murata Science and Education Foundation. C. G. and E. H. acknowledge support from the DFG program Fermi-NESt through Grant No. GE 602/4-1. Additionally, E. H. acknowledges funding by the DFG through CRC1143 (Project No. 247310070) and the Würzburg-Dresden Cluster of Excellence on Complexity and Topology in Quantum Matter—ct.qmat (EXC 2147, Project ID 390858490). Seunghyun Khim acknowledges support from the DFG through KH 387/1-1.
\end{acknowledgments}


%

\end{document}